\documentclass[conference,a4paper]{IEEEtran}
\usepackage[utf8]{inputenc} 
\usepackage[T1]{fontenc}
\usepackage{url}
\usepackage{ifthen}
\usepackage{cite}
\usepackage[cmex10]{amsmath}

\usepackage{graphicx}
\usepackage{amsfonts}
\usepackage{amssymb}
\usepackage{mathrsfs}
\usepackage{subfigure}
\usepackage{mdwmath}
\usepackage{mdwtab}
\usepackage{dsfont}
\usepackage[ruled,vlined,linesnumbered]{algorithm2e} 
\newtheorem{remark}{Remark}

\newenvironment{iarray}{\begin{IEEEeqnarray}{rCl}}{\end{IEEEeqnarray}\ignorespacesafterend}

\begin{document}

\title{%Index Policy for Status Update Scheduling with Stochastic Packet Arrivals
Distributed Policy Learning Based Random Access for Diversified QoS Requirements
}

\author{\IEEEauthorblockN{Zhiyuan Jiang$^{\ast \dagger}$, Sheng Zhou$^{\dagger}$,  Zhisheng Niu$^{\dagger}$, \IEEEmembership{Fellow,~IEEE}}
    \IEEEauthorblockA{$\ast$: Shanghai Institute for Advanced Communication and Data Science, Shanghai University, Shanghai 200444, China. \\
    	$\dagger$: Beijing National Research Center for Information Science and Technology, Tsinghua University, Beijing 100084, China.\\
    	Emails: zhiyjiang@foxmail.com, \{sheng.zhou, niuzhs\}@tsinghua.edu.cn 
      }}

    \maketitle

\begin{abstract}
Future wireless access networks need to support diversified quality of service (QoS) metrics required by various types of Internet-of-Things (IoT) devices, e.g., age of information (AoI) for status generating sources and ultra low latency for safety information in vehicular networks. In this paper, a novel inner-state driven random access (ISDA) framework  is proposed based on distributed policy learning, in particular a cross-entropy method. Conventional random access schemes, e.g., $p$-CSMA, assume state-less terminals, and thus assigning equal priorities to all. In ISDA, the inner-states of terminals are described by a time-varying state vector, and the transmission probabilities of terminals in the contention period are determined by their respective inner-states. Neural networks are leveraged to approximate the function mappings from inner-states to transmission probabilities, and an iterative approach is adopted to improve these mappings in a distributed manner. Experiment results show that ISDA can improve the QoS of heterogeneous terminals simultaneously compared to conventional CSMA schemes. 
\end{abstract}

%\begin{IEEEkeywords}
%Internet-of-Things, age-of-information, reinforcement learning, random access, quality-of-service
%\end{IEEEkeywords}

\section{Introduction}
The design principle for previous generations of wireless networks, e.g., 3G, 4G or even 5G Phase-1 cellular systems and Wi-Fi networks, was mostly throughput-focused. However, with the emergence of new applications in Internet-of-Things (IoT) systems, more diversified quality-of-service (QoS) requirements for future wireless networks need to be considered. Aligning with this goal, the 5G Phase-2 standard and foreseeable future wireless systems will try to address QoS of---in addition to enhanced mobile broadband---ultra-reliable and low latency communications (URLLC) and massive IoT access with heterogeneous types of terminals. Specifically, new IoT applications in e.g., vehicular networks, industrial IoT may have distinctive QoS requirements on wireless networks, while they co-exist in a single system. How to provide services with such \emph{diversified QoS guarantees} remains a critical challenge for future systems. 

Most existing approaches to address diverse QoS requirements focus on network slicing \cite{zhang17} techniques, whereas research on wireless multi-access protocol design is limited. The \emph{multiple access} (MA) scheme has been the research focus of wireless access networks for decades; in fact, the main distinguishing technology advances of 1G to 4G have been known for their multi-access schemes---from time/frequency division multiple access (T/FDMA), code-DMA (CDMA) to orthogonal frequency-DMA (OFDMA). On the other hand, Wi-Fi networks implement carrier sensing multiple access (CSMA) schemes wherein a contention-based random access principle is adopted. With their each pros and cons, the MA schemes of the cellular system and Wi-Fi system are evolving to be more and more converged, that is, future cellular systems adopt random access schemes to reduce the initial access delay in massive IoT systems and Wi-Fi systems also use scheduled access in certain scenarios to mitigate congestions. Nonetheless, almost all of the MA schemes in the literature are designed throughput-oriented, and not suitable for a system with diversified QoS requirements from heterogeneous types of terminals. 

\begin{figure}[!t]
\centering
\includegraphics[width=0.46\textwidth]{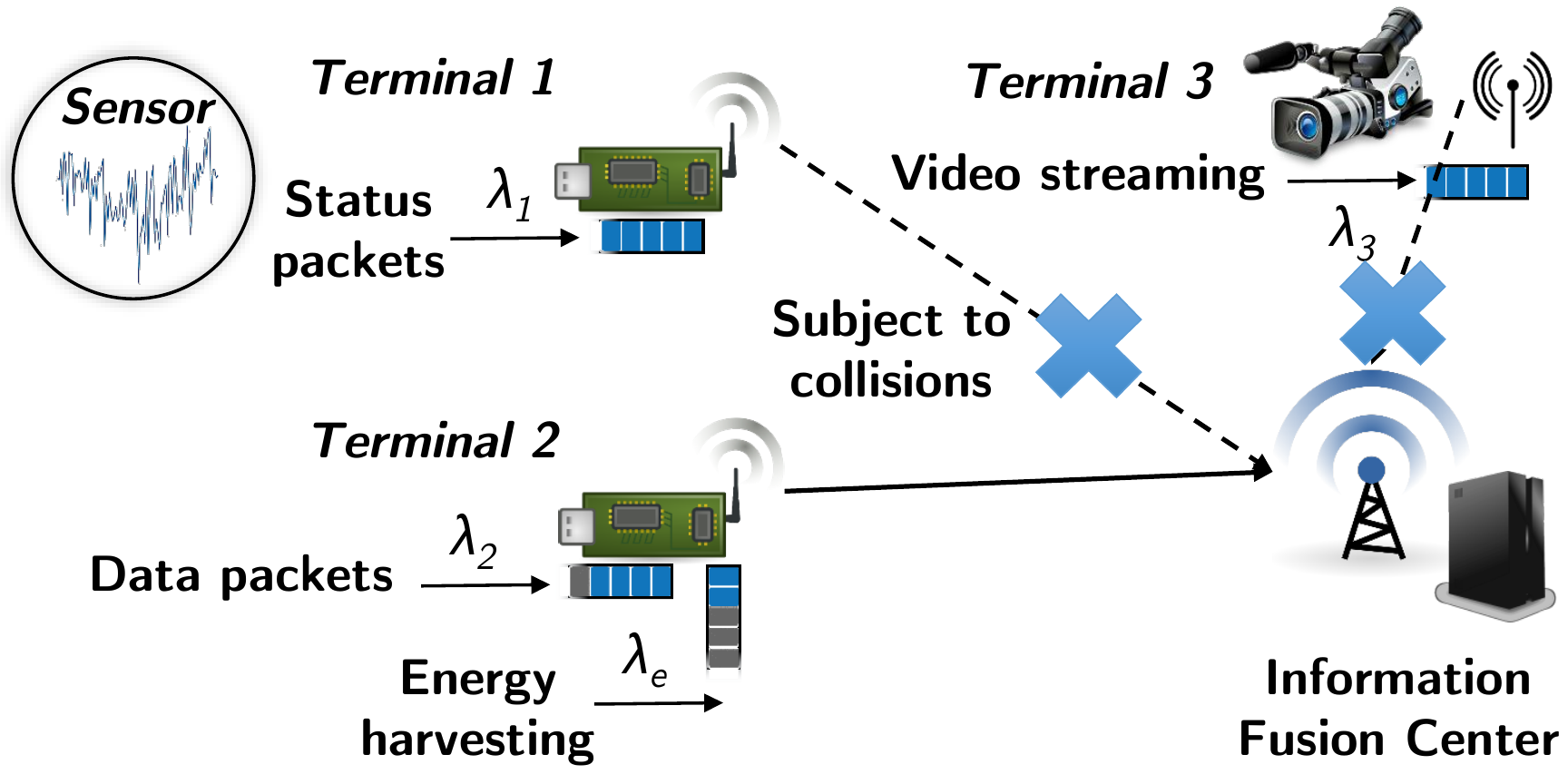}
\caption{A general wireless access network with diversified types of terminals.}
\label{fig_arch}
\end{figure}
In this paper, we generalize our previous work on age-of-information (AoI) optimization in multi-access networks \cite{jiang18_isit,jiang18_itc} to address the issue that how to deal with general, diversified QoS requirements. We develop a distributed policy learning based state-driven MA scheme wherein the QoS of terminals is optimized by defining the internal state of each terminal at every time slot and prioritizing the terminal with the most urgency in the MA process based on their individual internal states, i.e., \emph{inner-state-driven}. Since the abstraction from internal state to transmission urgency cannot be exhaustively derived for each application, a distributed policy learning based approach is leveraged whereby each terminal learns its transmission policy (a generalized $p$-CSMA scheme with parameters to be learned) and carries it out in a decentralized manner, in order to avoid prohibitively high signaling overhead in massive IoT systems. 

\subsection{Related Work}
Extensive work has been dedicated to optimizing different QoS objectives in wireless access networks for IoT applications. Aside from the conventional throughput QoS consideration \cite{neely08}, various types of latency metrics are among the main concerns. For example, the recently proposed AoI \cite{kaul12} can be regarded as a metric for information delay; a line of work \cite{jiang18_isit,jiang18_itc,kadota18,hsu18} has been focused on optimizing AoI in wireless multi-access (or broadcast) networks wherein Whittle's index was used to capture the transmission urgency of terminals. Conventional packet delay metric was widely investigated in the literature and various scheduling policies for multi-access networks have been proposed, most evidently in \cite{neely13}. Other QoS metrics such as inter-delivery time \cite{guo18} have also been discussed in cyber-physical systems. However, very few work has considered diversified QoS requirements.

This paper focuses on random access schemes, more specifically a generalized $p$-CSMA scheme is based upon. Conventional analysis and generalizations of CSMA schemes all focus on throughput optimization. Ref. \cite{gai11} derived the optimal transmission probability in $p$-CSMA schemes given greedy sources. Considering realistic traffic patterns, authors in \cite{ni12} proposed the queue-CSMA scheme which measures the transmission urgency of terminals by their queue lengths. For general QoS requirements, the transmission urgency is difficult to express explicitly, and hence we adopt a distributed policy learning approach \cite{ce} that can adapt to various QoS requirements automatically; as far as we know, this is the first work to do so.

\section{System Model and Problem Formulation}
\label{sec_sm}
The considered system model is shown in Fig. \ref{fig_arch}. There is an information fusion center, or cloud server, which is responsible for collecting information from heterogeneous terminals. The time is slotted and we assume the terminals are synchronized. Although synchronization is a critical issue in this setting, it is clearly out of the scope of the paper. The wireless access network is modeled by a multi-access channel, wherein the transmissions are subject to the protocol interference model, i.e., one transmission at a time. We assume reliable transmissions if no collision happens, although the learning-based approach can be easily generalized to unreliable transmissions. The transmission frame is structured as follows. 
\begin{figure}[!h]
	\centering
	\includegraphics[width=0.46\textwidth]{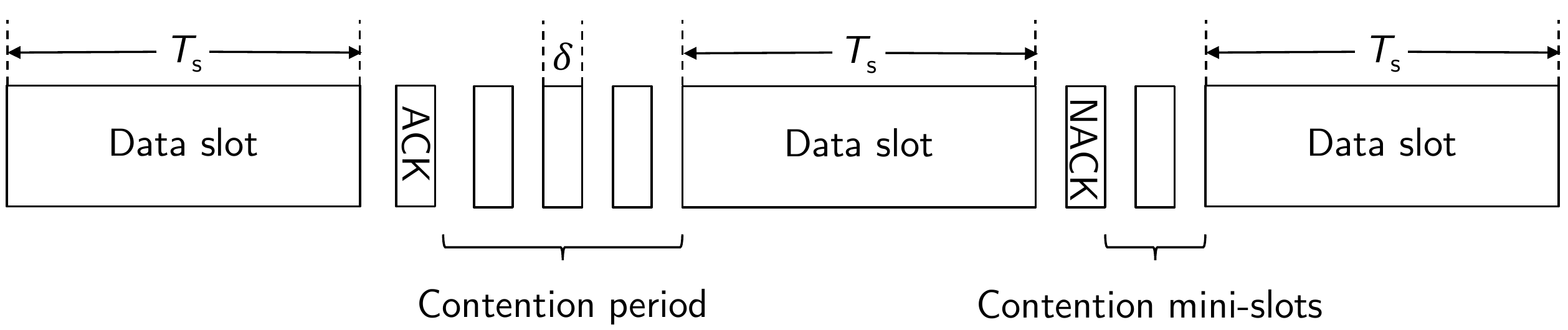}
	\label{fig_tx_frame}
\end{figure}

A data slot (with length $T_\textrm{s}$) is prefixed by several contention mini-slots (with length $\delta$). As in the $p$-CSMA scheme \cite{gai11}, terminal-$i$ transmit with a probability $p_i$ in each contention mini-slot; unlike the $p$-CSMA scheme but similar with the Q-CSMA scheme \cite{ni12}, the transmission probability $p_i$ is different among terminals, reflecting the transmission urgency. In the Q-CSMA scheme, $p_i$ is determined by the queue length of each terminal, which however only applies to throughput optimizations. After the contention period, the terminals that, based on their $p_i$, have won the contention transmit, subject to possible collisions. However, it is known that as long as the ratio $T_\textrm{s}/\delta$ is relatively large, the overhead of possible collisions and contention period is asymptotically approaching zero \cite{gai11}. After the data slot, the receiver feeds back an ACK/NACK to let the terminals know whether their transmissions succeed. 

Therefore, it is clear that the problem is how to select the transmission probability $p_i$ for each terminal such that the overall system performance is optimized. In addition, an important aspect is that each terminal should determine its transmission probability based solely on its own inner-state (defined later), i.e., making the scheduling implementation decentralized. This is especially important in future massive IoT systems, in order to avoid the prohibitive high signaling overhead otherwise. 

\section{Inner-State-Driven Random Access for Diversified QoS Requirements}
The decentralized selection of transmission probabilities is hence the problem of interest. For single QoS requirement, e.g., throughput and AoI optimization, this problem has been addressed separately; Ref. \cite{jiang18_itc} derived Whittle's index and mapped it to transmission probability based on a heuristic threshold-based policy for AoI optimization; Ref \cite{ni12} optimized throughput by designing a queue-based CSMA scheme based on the Glauber dynamics. However, there is significant difficulty in designing the random access scheme for diversified QoS requirements, due to the fact that it is almost impossible to theoretically derive the index-type transmission urgency indicator for each QoS requirement, and additionally it is unclear how to determine the transmission priority when different types of QoS are jointly considered. 

For concreteness, we consider the following three types of QoS requirements in this paper, representing the typical application scenarios in future IoT systems. Note that more QoS types can be readily plugged into the learning-based approach.

1) AoI minimization for status update systems. The AoI is formally defined as 
\begin{equation}
\underline{\textrm{AoI at time }t} \triangleq t - \mu(t),
\end{equation} 
where $\mu(t)$ is the generation time of the most up-to-date packet at the receiver side until time $t$. AoI is particular useful to characterize the data freshness from a destination perspective, and hence useful in control systems wherein the control action stringently depends on timely status information from sensors, whose data are collected through a wireless network. At each terminal, the status generation is determined by the status variation, e.g., temperature variation, and a sampling scheme; in this work, we assume the status packets arrive randomly at the terminal. When a packet first arrives, its AoI is one and increases afterwards. We adopt the optimal packet management at terminal that only keeping the freshest packet \cite{jiang18_itc}. The inner-state of terminal-$i$ for AoI optimization is defined as
\begin{equation}
\label{s_aoi}
S_{\textrm{AoI},i} = (a_i,h_i),
\end{equation}
where $a_i$ is the age of packet at the terminal queue, and $h_i$ denotes the AoI at destination which the terminal knows by keeping track of the receiver feedbacks. The AoI evolution can be formulated by
\begin{iarray}
	\label{evo}
	&& h_{i}(t+1) =  h_{i}(t) + 1 - x_{i}(t) (h_i(t)-a_i(t)), 
\end{iarray}
where the binary variable $x_{i}(t)$ denotes whether the update is successful. The QoS objective for this kind of terminal is minimize the long-time average AoI, which is defined as
\begin{equation}
\label{AoI}
\bar{V}_\textrm{AoI} \triangleq \limsup_{\tau \to \infty}  \frac{1}{\tau N}\sum_{t=1}^\tau \sum_{n=1}^N  \mathbb{E}[h_{n}(t)],
\end{equation}

2) Inter-delivery time minimization with energy harvesting transmitters. In future IoT systems, it is envisioned that transmitters with energy harvesting capabilities are ubiquitous. This poses a challenge in the random access scheme design that not only the transmission urgency of each terminal should be considered as in the AoI minimization case, the transmission capability should also be taken into account. The energy harvesting transmitters are modeled by energy buffers with size of $B$. The energy packets arrive randomly at the energy buffers, and each energy packet can provide the energy to transmit one data packet in one data slot; note that we have neglected the energy consumption of contention period in this work. The terminal cannot transmit when there is no packet in its energy buffer, and packet arrivals when the buffer is full are dropped due to battery capacity limitations. In addition, we consider in this case the inter-delivery time \cite{guo18} for the terminal, representing the time gap between successive packet deliveries, which is also equivalent to AoI with packet arrival rates of one. Based on the aforementioned energy harvesting transmitters and inter-delivery time consideration, the inner-state of terminal-$i$ is defined as
\begin{equation}
\label{s_idt}
S_{\textrm{IDT-EH},i} = (d_i,e_i),
\end{equation}
where $d_i$ denotes the current time elapsed since the last packet delivery, and $e_i$ denotes the energy buffer queue length. The inter-delivery time evolution is
\begin{iarray}
	\label{idt_evo}
	&& d_{i}(t+1) =  d_{i}(t) + 1 - x_{i}(t) d_i(t), 
\end{iarray}
where the binary variable $x_{i}(t)$ denotes whether the update is successful. The QoS objective is to minimize the average long-time average inter-delivery time with the energy constraint.

3) Throughput maximization. Finally, we also consider terminals with the objective to maximize their throughput. For these terminals the inner-state is the traditional queue length:
\begin{equation}
\label{s_r}
S_{\textrm{R},i} = q_i,
\end{equation}
and its evolution is formulated by
\begin{iarray}
	\label{q_evo}
	&& q_{i}(t+1) =  \max(q_{i}(t) + a_i(t) - x_{i}(t),0), 
\end{iarray}
where $a_i(t)$ is binary representing whether there is a packet arrival at time $t$. It is well known that scheduling based on the queue length, i.e., the max-weight scheduling, is throughput optimal \cite{tas92} and therefore is used to consider throughput maximization.

\subsection{Inner-State-Driven Random Access}
The three types of QoS requirements described previously are not restrictive, and in general, as long as the inner-state of the terminal can be described as a real-valued vector, i.e.,
\begin{equation}
\label{s_general}
\underline{\textrm{Inner state of terminal-}i}  \triangleq S_{i} \in \mathcal{S}_i
\end{equation}
as in \eqref{s_aoi}\eqref{s_idt}\eqref{s_r}, where $\mathcal{S}_i$ is the state space of terminal-$i$, the following framework applies. 
\begin{figure}[!t]
	\centering
	\includegraphics[width=0.46\textwidth]{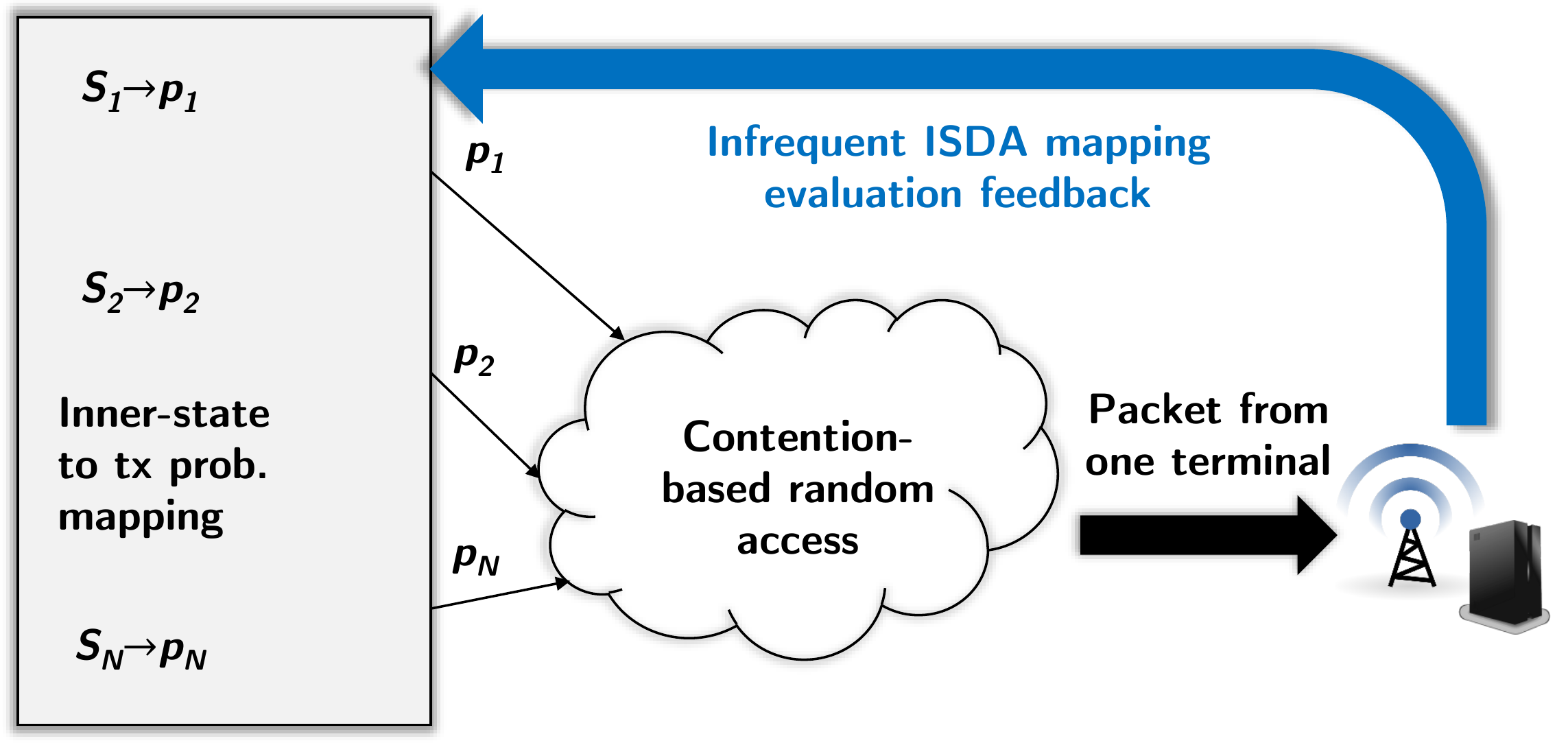}
	\caption{Framework of ISDA wherein the inner-state is mapped to transmission probability for each terminal.}
	\label{fig_isda}
\end{figure}

We describe the inner-state-driven random access (ISDA) framework in Fig. \ref{fig_isda}. For each terminal, its inner-state $S_i$ is mapped to a transmission probability $p_i$, where the mapping function is to be discussed in the next section. After each terminal decides its transmission probability thereby, a contention-based random access protocol is implemented; the frame structure is described in Section \ref{fig_tx_frame}. Only the packet from one terminal can be transmitted in each time slot, with probability of collision, to the information fusion center. The fusion center collects packets and feeds back evaluations of the mapping functions implemented at each terminal, and the terminals iteratively improve their respective mapping functions based on the feedback, supposing that the mapping functions are constructed in such a way that can be improved, such as a parameterized neural network. 

The proposed ISDA framework has the advantage of self-optimizing networks, i.e., it can not only adapt to diversified QoS requirements given universal mapping functions construction which is discussed in the next section, it also adapts to environmental changes such as channel error changes or the dynamics of terminals, as long as the iterative improvements converge.  

\section{Distributed Policy Learning Based Mapping Function Optimization}
The design of the inner-state to transmission probability mapping functions at terminals faces two critical challenges. First, diverse QoS and inner-state descriptions pose significant difficulty to derive the mapping functions explicitly for every scenario, and also quite challenging to be adaptive to environmental changes. Secondly, the transmission protocol should be carried out in a decentralized manner to save signaling overhead, thus making it challenging to optimize the mapping functions due to lack of state information of other terminals. 

To tackle these challenges, we adopt a distributed policy learning based approach with infrequent feedback from the receiver as shown in Fig. \ref{fig_isda}. Before presenting the main algorithm, some preliminaries are formulated to better understand the problem. By leveraging the learning framework for the sake of better adaptiveness, the problem is, by nature, a reinforcement learning problem due to lack of supervised expert providing the optimal action; in particular, the problem can be seen as a \emph{partially observable identical payoff stochastic game} (POIPSG) \cite{pes00} which is used to model a problem wherein multiple agents learn simultaneously with a single objective and observations of only local state information. The notion of game represents the involvement of multiple agents, not necessarily meaning playing against each other. 

For brevity, we neglect the details of denotations in POIPSG, and simply denote the entire state as $(S_1, S_2, \cdots ,S_N)$ and the observation for terminal-$i$ is $S_i$; the action of each terminal is whether to transmit, i.e., $a_i\in \mathcal{A}=\{0,1\}$, with the transmission probability $p_i $; we further denote $T:\mathcal{S}_1 \times \cdots \mathcal{S}_N \times \mathcal{A}^N\rightarrow \mathcal{P}(\mathcal{S}_1 \times \cdots \mathcal{S}_N)$ as the system transition mapping. The reward for each terminal can only be observable at the receiver side and hence is decided by the receiver subject to certain credit assignment algorithm; this is different with traditional POIPSG settings wherein the reward signal is assumed to be known to all agents, thus posing a challenge which is to be addressed in this section later. The objective is set to optimize the weighted average QoS objectives of all the terminals, i.e.,
\begin{equation}
\label{w_qos}
\bar{V} = \sum_{n=1}^N \omega_n \bar{V}_{\textrm{QoS},n},
\end{equation}
where $\omega_n$ is the weight, and $\bar{V}_{\textrm{QoS},n}$ denotes the QoS objective function of terminal-$n$, e.g., in \eqref{AoI}. We consider the set of stochastic reactive policies wherein the action of the terminal only depends on the current inner-state, and the dependence can be probabilistic. 

In general, there are two categories of reinforcement learning algorithms, i.e., policy optimization methods, e.g., cross-entropy methods and policy gradient methods, and value-function based methods, e.g., Q-learning and SARSA methods \cite{sch17}. It is well known that, although value-function based methods achieved some tremendous results for completely observable case, they are problematic when applied to partially observable problems, as in our considered problem, due to the fact that the value functions are insufficient to describe the value of each state given partially observable states. In this regard, we adopt a policy optimization method, in particular a cross-entropy based approach \cite{szi06}, to accomplish the task. The detailed algorithm is described as follows.

\subsection{Episodic Setting}
An episodic setting is considered to facilitate the interaction between terminals and the receiver, which will be discussed in the following subsections. Each episode lasts $T_\textrm{epi}$ data slots. The receiver maintains the records of the weighted QoS metric \eqref{w_qos} of each episode. After $M_\textrm{fb}$ episodes, the receiver evaluates the terminals' policies by collecting the episodic QoS metrics during the $M_\textrm{fb}$ episodes, and feeds back policy improvement directions to each terminal, wherein the detailed feedback information is given in the following subsections. We denote the $t$-th episode $(1 \le t \le T_\textrm{epi})$ in the $m$-th evaluation iteration $(1 \le m \le M_\textrm{fb})$ in the subscript of all denotations.

\subsection{Feedback from Receiver}
The receiver should send feedback to terminals to reflect the evaluations of their policies. In this work, we adopt the cross-entropy method whereby the receiver obtains the transmissions of $M_\textrm{fb}$ episodes and selects the best $\lfloor\rho M_\textrm{fb} \rfloor$ episodes, where $\rho$ is a hyper-parameter denoting the selection ratio. Then the receiver feeds back the index set of the selected episodes, which is denoted by $I_m$.

\begin{remark}
	Note that the overall performance of all terminals is selected as the evaluation metric by our design. An alternative is to use the performance (AoI, queue length and etc.) of each terminal to evaluate them respectively, making the problem a multi-agent stochastic game with conflicting goals. Usually that would result in divergence and hence a common goal is selected. However, the proposed approach with a common goal faces the problem of credit assignment among terminals, which is identified as the main problem for multi-agent learning. We have not addressed this problem completely and the current approach seems to work in the considered scenario.
\end{remark}
\subsection{Action Selection at Terminals}
We adopt neural networks (NNs) as approximation function mappings from terminal inner-states, e.g., $S_i$ as input, to actions. The output of the NN for terminal-$i$ is the transmission probability of terminal-$i$ in the contention period. Denote the parameter vector of the NN for terminal-$i$ as $\boldsymbol{w}_i$ with length $L_i$. In the $t$-th episode, $\boldsymbol{w}_{i,t,m}$ is generated by drawing a sample from a Gaussian distribution $\boldsymbol{w}_{i,t,m} \sim \mathcal{N}(\boldsymbol{\mu}_{t,m},\boldsymbol{\Sigma}_{t,m})$, where $\boldsymbol{\Sigma}_{t,m}=\operatorname{diag}(\sigma_{t,m,1}^2,...,\sigma_{t,m,L_i}^2)$. The policy in the $t$-th episode is determined by using $\boldsymbol{w}_{i,t,m}$. In the next episode-$(t+1)$ of the same evaluation iteration, the parameters are re-drawn from the same distribution. 

After obtaining the feedback from the receiver, each terminal updates its NN parameters generating distributions by 
\begin{equation}
\boldsymbol{\mu}_{t,m+1} = \frac{\sum_{t \in I_m}\boldsymbol{w}_{i,t,m}}{|I_m|},
\end{equation}
where $|\cdot|$ denotes the cardinality of the set, and 
\begin{equation}
\sigma_{t,m+1,1}^2 = \frac{\sum_{t \in I_m}(\boldsymbol{w}_{i,t,m}-\boldsymbol{\mu}_{t,m+1})^\mathsf{T}(\boldsymbol{w}_{i,t,m}-\boldsymbol{\mu}_{t,m+1})}{|I_m|}.
\end{equation}
In other words, a Gaussian distribution is fit to the selected episode and is used to generate NN parameters in the next evaluation iteration.

\begin{figure}[!t]
	\centering
	\includegraphics[width=0.46\textwidth]{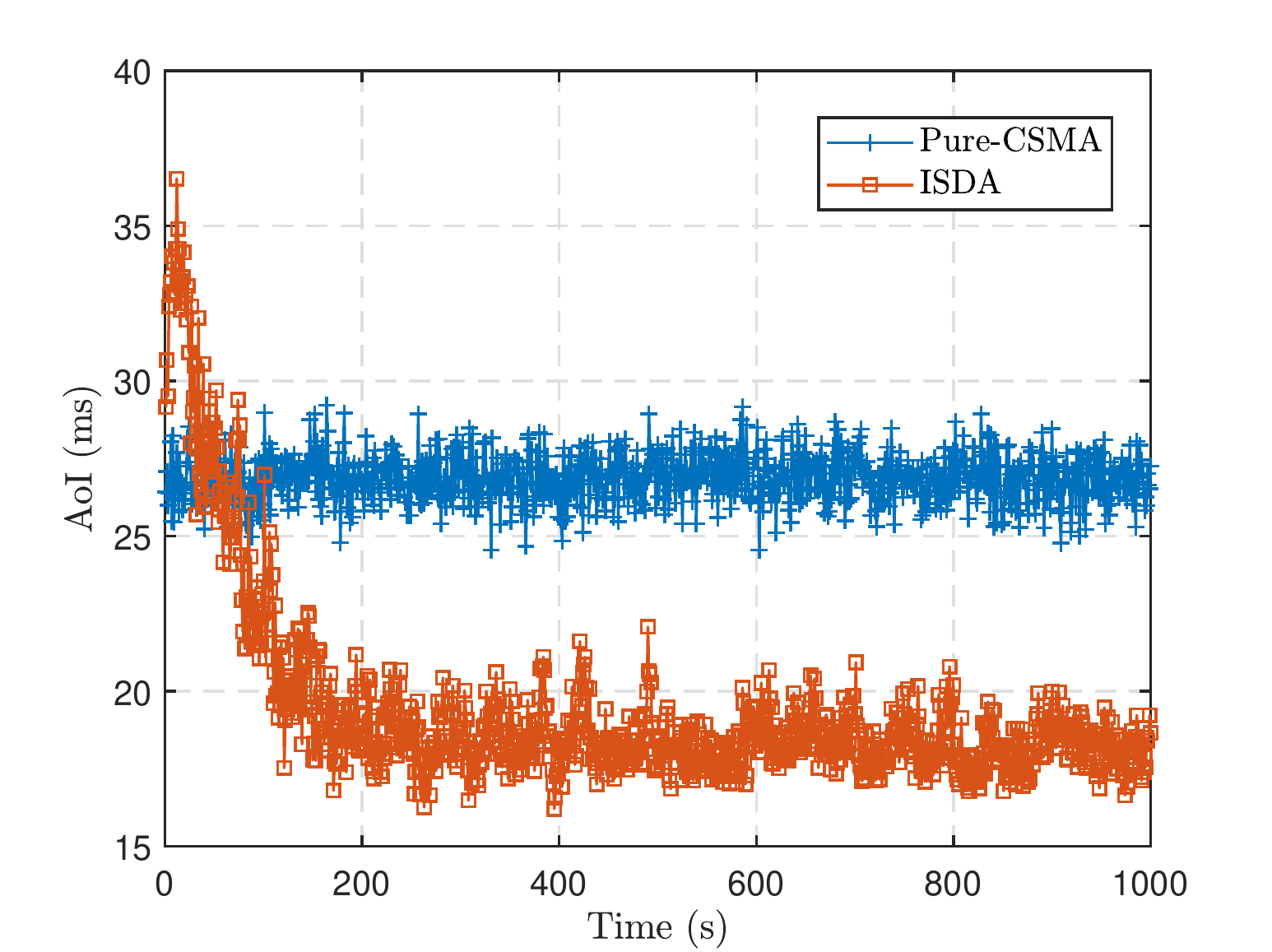}
	\caption{The AoI comparisons between ISDA and pure-CSMA in the $3$-terminal case.}
	\label{fig_aoi}
\end{figure}
\begin{figure}[!t]
	\centering
	\includegraphics[width=0.46\textwidth]{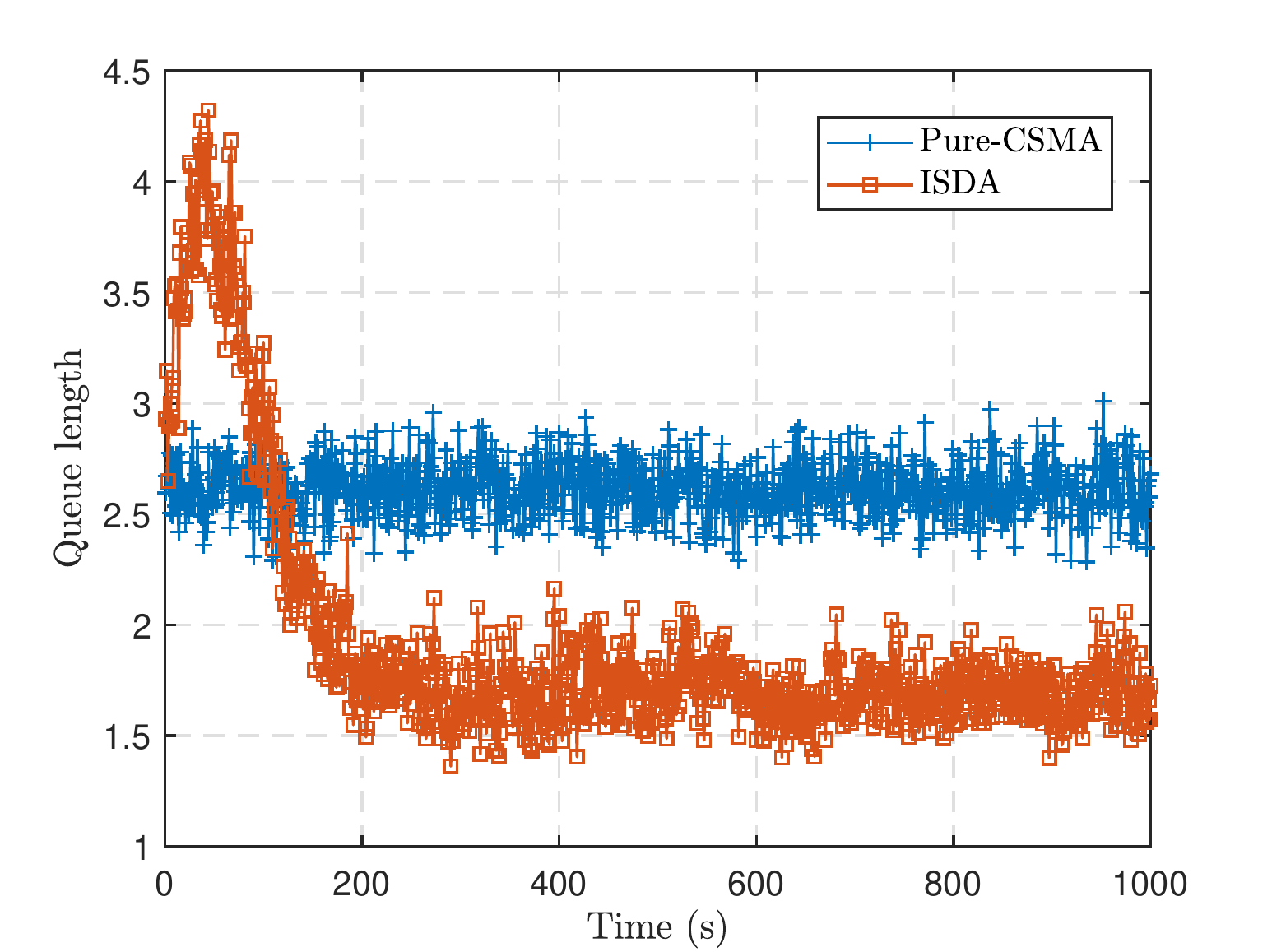}
	\caption{The queue-length comparisons between ISDA and pure-CSMA in the $3$-terminal case.}
	\label{fig_queue}
\end{figure}
\begin{figure}[!t]
	\centering
	\includegraphics[width=0.46\textwidth]{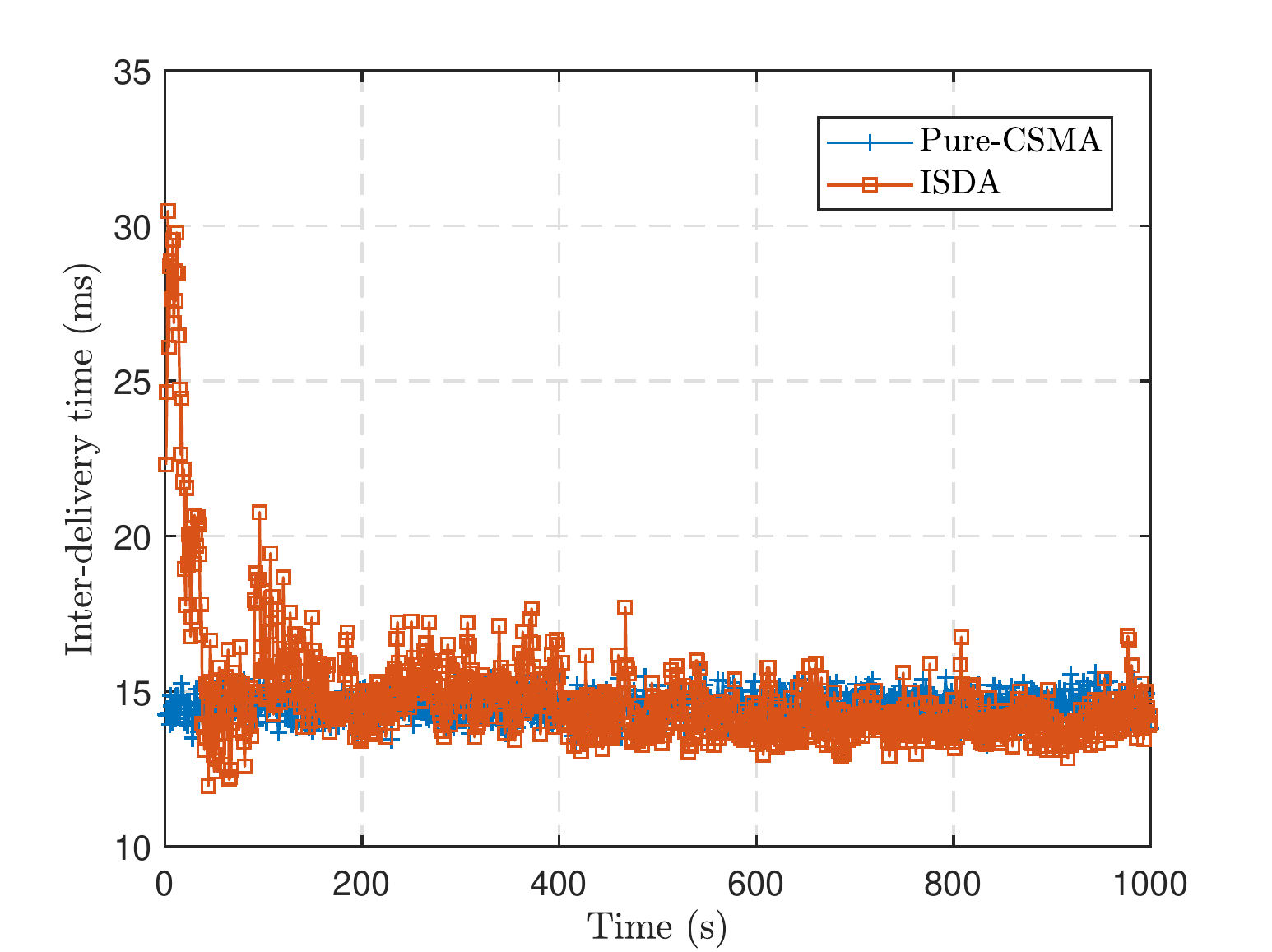}
	\caption{The inter-deliver time comparisons between ISDA and pure-CSMA in the $3$-terminal case.}
	\label{fig_idt}
\end{figure}
\section{Experiments}
\label{sec_nr}
We test our proposed ISDA framework by a computer simulation, with a simplified system setting wherein $3$ terminals are considered. The first one is to optimize its AoI with status packets arriving based on a discrete Bernoulli process with mean rate of $0.1$ packets/ms; the optimal packet management for AoI optimization is considered wherein older packets are dropped whenever there is a new packet arrival. The second is a video streaming terminal whose packets arrive based on a discrete Bernoulli process with mean rate of $0.1$ packets/ms and they are served by a first-come-first-served (FCFS) discipline; the performance metric is the average queue length, which corresponds to the average queuing delay based on Little's Theorem, and more importantly, it is well-known that queue-weighted scheduling optimizes throughput \cite{tas92}. The third is a energy harvesting terminal that tries to minimize its inter-delivery time; the energy buffer size is assumed to be one and the energy arrivals obey a discrete Bernoulli process with arrival rate $0.2$ packets/ms. 

The duration of a data slot is assumed to be $T_\textrm{s}=1$~ms, in line with the current LTE subframe length. Each contention mini-slot is assumed to be $\delta=0.25$~ms. In the cross-entropy based ISDA scheme, the NN consists of one hidden layer with $5$ neurons; the activation function of the hidden layer is ReLu, and that of the output layer is Softmax. The number of data slots in each episode is set to be $ T_\textrm{epi}=100$, and the number of episodes for evaluation at the receiver is $M_\textrm{fb}=100$. We select the best $10\%$ episodes for feedback, i.e., $\rho=0.1$. In addition, we adopt the cross-entropy method with decreasing noise which is a well-known method to avoid premature convergence to a sub-optimum, that is, the noise variance of the generating Gaussian distribution is re-written as
\begin{iarray}
\sigma_{t,m+1,1}^2 &=& \frac{\sum_{t \in I_m}(\boldsymbol{w}_{i,t,m}-\boldsymbol{\mu}_{t,m+1})^\mathsf{T}(\boldsymbol{w}_{i,t,m}-\boldsymbol{\mu}_{t,m+1})}{|I_m|} \nonumber\\
&& + z_{m+1},
\end{iarray}
where $z_{m+1} = z_0/(m+1)$, and $z_0$ is an initial noise variance. We set the terminal weights as $\omega_1=\omega_2=\omega_3=1$. 

\begin{table}[!t]
	\caption{Performance comparisons between ISDA and pure-CSMA}
	\label{tab}
	\centering
	\begin{tabular}{| c | c | c | c |}
		\hline
		& AoI & Queue-length & Inter-delivery time  \\
		\hline
		Pure-CSMA & $26.9$~ms & $2.28$ & $15.5$~ms \\
		\hline
		ISDA & $20.7$~ms & $1.41$ & $12.6$~ms \\
		\hline
	\end{tabular}
\end{table}
In Fig. \ref{fig_aoi}-\ref{fig_idt}, we compare our proposed ISDA framework with the pure-CSMA scheme. In pure-CSMA scheme, every terminal gets equal probability to transmit, regardless of their inner-state variation. The time-average performance after ISDA converges is shown in Table \ref{tab}. It is observed that the performances of all terminals with different QoS requirements can be improved by ISDA, due to the fact that ISDA takes advantage of the inner-states of each terminal. In the figures, it is also observed that the convergence time is about $200$~s; this is reasonably acceptable when the terminals in IoT systems are relatively static, and hence it is affordable for the network to take several minutes to self-optimize. 

\begin{figure}[!t]
	\centering
	\includegraphics[width=0.46\textwidth]{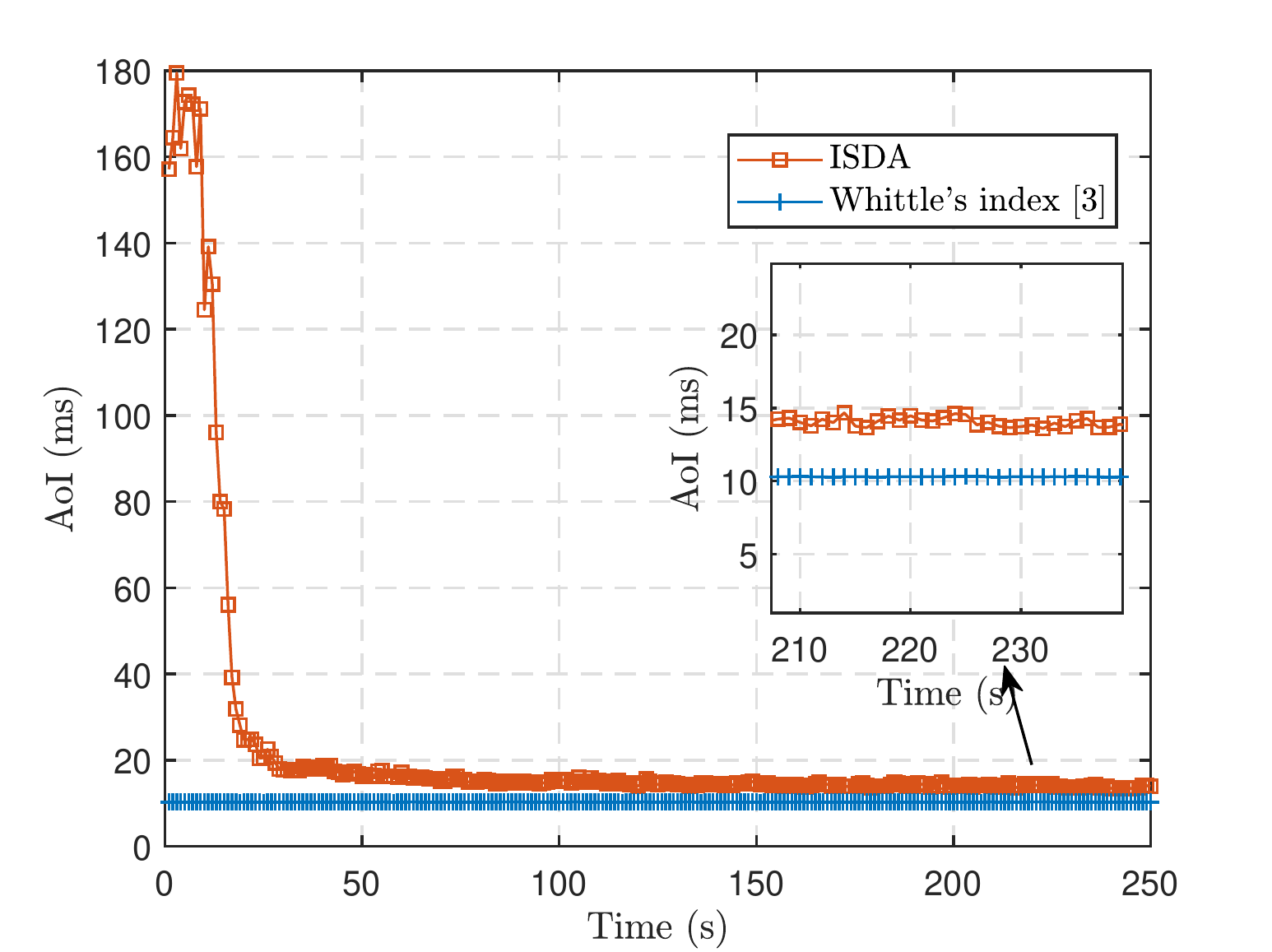}
	\caption{The AoI comparisons between ISDA and the (near-)optimal solution in the $3$-terminal case.}
	\label{fig_aoi_opt}
\end{figure}

\subsection{Comparisons with Near-Optimal Solution}
In general scenarios, thee is no known optimal solution. However, it is important to understand the extend of sub-optimality of our proposed ISDA, and thus we consider a scenario wherein $3$ terminals all try to optimize their AoI. The reason we adopt this scenario is that in our previous work \cite{jiang18_itc}, we have theoretically derived the (near-)optimal random access scheme in this setting based on the Whittle's index approach. In particular, the Whittle's index is given by $m(a_n,b_n,\lambda_n)=\frac{1}{2} x_n^2 + \left(\frac{1}{\lambda_n}-\frac{1}{2}\right) x_n$, if $b_n > \frac{\lambda_n}{2}(a_n^2-a_n)+a_n$; and $\frac{b_n}{\lambda_n}$ otherwise, where $x_n\triangleq \frac{b_n + \frac{a_n(a_n-1)}{2}\lambda_n}{1-\lambda_n+a_n \lambda_n}$, the age of the packet at buffer of terminal-$n$ is denoted by $a_n$, the current AoI of terminal-$n$ is denoted by $h_n$ and $b_n \triangleq h_n - a_n$, the packet arrival probability is $\lambda_n$. It is well-known that the Whittle's index approach is asymptotically optimal The transmission probability is chosen by mapping the index to a scaler value based on a threshold-based function mapping. It is observed from Fig. \ref{fig_aoi_opt} that the ISDA performance is close to the optimum achieved by \cite{jiang18_itc} after convergence. Also the convergence time is much smaller compared with the previous case with heterogeneous types of terminals. 

\section{Conclusions and Future Work}
\label{sec_con}
In this paper, we propose an inner-state driven random access framework for future wireless access networks with heterogeneous QoS requirements. Based on distributed policy learning, in particular a cross-entropy based approach, the transmission opportunity is scheduled among terminals by mapping the inner-states of each terminal to a transmission probability and letting them participate in the random access contention. Experiment results show that, in a simplified case with $3$ heterogeneous terminals, the QoS of all terminals can be improved simultaneously by ISDA compared to conventional CSMA schemes. Furthermore, by comparing with our previous work, it is shown that ISDA performs close to optimum when only AoI is considered. 

The ISDA framework is very versatile in the sense that it is applicable as long as the state of each terminal can be described by a real-valued vector. However, we note that there are several open issues that need to be addressed in future work. The policy optimization method needs further improvement, regarding convergence rate and performance. The scalability problem is also significant considering the application scenarios.

\section*{Acknowledgement}
This work is sponsored in part by the Nature Science Foundation of China (No. 61861136003, No. 61871254, No. 91638204, No. 61571265, No. 61621091), National Key R\&D Program of China 2018YFB0105005, and Hitachi Ltd.

\bibliography{sm}
\bibliographystyle{IEEEtran}
\end{document}